# Conversion of magnetic freedoms into atomic configurational freedoms within the Cluster Variation Method

Ryo Yamada* and Tetsuo Mohri

Institute for Materials Research, Tohoku University, Sendai 980-8577, Japan

* Corresponding author. E-mail address: ryamada@imr.tohoku.ac.jp

## Abstract

The continuous displacement cluster variation method (CDCVM) has introduced local atomic displacements into the theoretical framework of the cluster variation method (CVM) by viewing an atom displaced from a Bravais lattice point as a particular atomic species located at the lattice point. This idea of conversion from a freedom of local displacements into configurational freedom is extended in this paper to magnetic freedoms. Various magnitudes of local magnetic moments are considered, as well as two spin directions, on up-spins and down-spins. The approach is applied to pure Ni and its Curie temperature is explored with the entropy formula of the tetrahedron approximation in the CVM, using the first-nearest-neighbor pair interaction energies extracted from the total energies of various spin configurations, which are estimated from electronic-structure calculations.

## 1. Introduction

The cluster variation method (CVM)[1] is a powerful theoretical tool for calculating free energies on an atomistic scale. It has been used to derive a variety of thermodynamic properties of alloy systems, such as phase diagrams,[2] spinodal ordering temperature,[3, 4] Curie temperature,[5, 6] specific heat capacity,[6] and coefficient of thermal expansion.[7, 8] In general, the bigger the basic cluster (i.e., the largest cluster considered in the free-energy formula) employed, the more accurate the result that can be obtained.

Conventional CVM is formulated on a lattice that maintains Bravais symmetry over all lattice points; consequently, it permits only a uniform lattice expansion or contraction, and local lattice relaxation is not taken into account. As a result, a disordered phase in which atoms of different sizes have a greater chance of encountering one another tends to be understabilized in comparison with an ordered phase. This induces an overestimation of order–disorder transition temperatures. The introduction of a local lattice displacement, however, disturbs the original crystal symmetry, so that the entropy formula of conventional CVM is no longer justified. To circumvent this inconvenience, Kikuchi devised the continuous displacement cluster variation method (CDCVM).[9]

In the CDCVM, additional points are introduced around a Bravais lattice point, and an atom is allowed to displace to one of these points. These additional points are termed 'quasi-lattice points'. For each quasi-lattice point, a different atomic species is assigned, and an atom displaced to a quasi-lattice point is regarded as a particular atomic species (assigned to the quasi-lattice point) located at a Bravais lattice point. Thus, the freedom of atomic displacement is replaced by a configurational freedom of a multicomponent alloy on a rigid or uniformly deformable lattice. Because a vast number of configurational variables are involved in CDCVM, its use has been limited to a few tractable cases, such as two-dimensional lattices.[10, 11] However, the application of CDCVM is expected to be enhanced with the development of more powerful computers.

In addition to its significance in terms of improving the accuracy of calculated results, CDCVM should be considered as a means of converting additional freedom into configurational freedom. This idea of conversion of freedom is not limited to a local displacement, and it can be applied to collective displacements leading to a phase transition. In fact, the cubic–tetragonal transition of $ZrO_2$ has been studied by regarding the upward- or downward-shifted oxygen atom as a different species located on a cubic lattice point.[12] Within such a treatment, the cubic–tetragonal transition can be viewed as an order–disorder transition on a cubic lattice; i.e., a displacive transition can be investigated within the realm of replacive transition.[13]

In a given lattice, various freedoms coexist or compete with one another, giving rise to various alloy properties as well as versatile transition phenomena. CDCVM opens up a new challenge for studying such freedoms within the well-tuned configurational thermodynamics of CVM. The present study focuses on a freedom of spin configurations that leads to a magnetic transition. The Curie temperature, in the present study, is defined as the temperature at which the number of up-spins equals the number of down-spins. Although the Curie temperature has been calculated by CVM,[5, 6] most studies have assumed invariant magnitudes of local magnetic moments, where up- and down-spins are treated as A and B atoms in an A–B binary alloy system. In this paper, not only up- and down-spins but also various magnitudes of magnetic moments are incorporated by viewing them as different atomic species, and the Curie temperature of pure Ni is explored using pair interaction energies extracted from total energies calculated from electronic-structure calculations.

This paper is organized as follows. Section 2 reviews conditions for calculating total energies and the theoretical procedures for calculating a pair interaction energy, an entropy, and a free energy. The results are then presented and discussed in Section 3.

## 2. Calculation procedure

### 2.1. Total energy calculation

The electronic-structure total-energy calculations are performed by means of the projector augmented-wave method,[14] as implemented in the Vienna Ab Initio Simulation Package (VASP) code. The VASP code with Perdew–Burke–Ernzerhof parameterization[15] is employed in the generalized gradient approximation.[16] The total energies of pure Ni with a face-centered cubic (fcc) structure in its nonmagnetic (nm), ferromagnetic (fm), and antiferromagnetic (af) states are calculated; for the nonmagnetic state, a non-spin-polarization calculation is conducted. The spin alignment of each phase is shown in Figs. 1(a)–(d). Among various antiferromagnetic configurations, two different states, where the spin configuration is reversed alternatively in the <001> or <111> direction, are considered here. These antiferromagnetic configurations are called 'type-I' and 'type-II', respectively, and correspond to $L1_0$ and $L1_1$ ordered phases in an A–B binary alloy. Supercells in the present calculations contain four atoms for nm, fm, and af-type-I, and eight atoms for af-type-II, and the plane wave cut-off is set at 810 eV. The number of *k*-points of 11X11X11 is used for the integration over the Brillouin zone.

### 2.2. Pair interaction energy

Although magnetic (or spin) interaction energies extend over long-range pairs and multi-body clusters, the contribution of the first-nearest-neighbor pair is considered to be mostly dominant.[17] Thus, considering this pair exclusively is the simplest approach to take. In the present calculation, a Lennard–Jones-type potential is employed to represent a pair interaction energy, which is given by the sum of repulsive and attractive terms:

$$e_{ij}(r) = e_{ij}^{0}\left\{\left(\frac{r_{ij}}{r}\right)^{m} - \frac{m}{n}\left(\frac{r_{ij}}{r}\right)^{n}\right\}, \tag{1}$$

where $i$ and $j$ indicate up-spin or down-spin; $m$, $n$, $e_{ij}^{0}$ and $r_{ij}$ are the fitting parameters; and $r$ is an interatomic distance.

To extract the pair energies between up-spins and up-spins (*u*-*u*), $e_{\uparrow\uparrow}$, and those between up-spins and down-spins (*u*-*d*), $e_{\uparrow\downarrow}$, the total energies calculated in Sec. 2.1 are used. First, a procedure to extract $e_{\uparrow\uparrow}^{\mathrm{fm}}$ from the total energy of the ferromagnetic state, $E_{\mathrm{fm}}$, is demonstrated. The total energy in a spin system, $E$, is described in terms of pair

interaction energies as

$$E = \frac{1}{2} NZ \sum_{i,j} e_{ij} y_{ij} \,, \tag{2}$$

where $Z$ is the coordination number, $N$ is the total number of lattice points, and $y_{ij}$ is the pair cluster probability of finding an *i–j* spin configuration. Because a ferromagnetic state is composed exclusively of atoms with up-spin, the total energy per lattice point can be expressed as

$$E_{\mathrm{fm}} = \frac{1}{2} \cdot 12 \cdot e_{\uparrow\uparrow}^{\mathrm{fm}} \,, \tag{3}$$

where $Z = 12$ for fcc structures is assigned. Therefore, $e_{\uparrow\uparrow}^{\mathrm{fm}}$ can be written as

$$e_{\uparrow\uparrow}^{\mathrm{fm}} = \frac{1}{6} \cdot E_{\mathrm{fm}} \,. \tag{4}$$

Note that the nonmagnetic pair interaction energy (*n-n*) is obtained in the same way, by replacing $E_{\mathrm{fm}}$ with $E_{\mathrm{nm}}$ (where $E_{\mathrm{nm}}$ is a nonmagnetic total energy). Secondly, $e_{\uparrow\downarrow}^{\mathrm{af}}$ is derived using an antiferromagnetic total energy, $E_{\mathrm{af}}$, and pair energies between up-spins and up-spins, $e_{\uparrow\uparrow}^{\mathrm{af}}$ . The way in which $e_{\uparrow\downarrow}^{\mathrm{af}}$ is extracted depends on which antiferromagnetic state is employed (type-I or type-II). The total energies of the two antiferromagnetic states within the first-nearest-neighbor pair interaction model become

$$E_{\mathrm{af,typeI}} = \frac{1}{2} \cdot 12 \cdot \left( \frac{2}{6} \cdot e_{\uparrow\uparrow}^{\mathrm{af,typeI}} + \frac{4}{6} \cdot e_{\uparrow\downarrow}^{\mathrm{af,typeI}} \right) , \tag{5}$$

$$E_{af,\mathrm{typeII}} = \frac{1}{2} \cdot 12 \cdot \left( \frac{1}{2} \cdot e_{\uparrow\uparrow}^{\mathrm{af,typeII}} + \frac{1}{2} \cdot e_{\uparrow\downarrow}^{\mathrm{af,typeII}} \right) , \tag{6}$$

and $e_{\uparrow\downarrow}^{\mathrm{af}}$ for type-I and type-II is given as

$$e_{\uparrow\downarrow}^{\mathrm{af,typeI}} = \frac{1}{4} \cdot E_{\mathrm{af,typeI}} - \frac{1}{2} \cdot e_{\uparrow\uparrow}^{\mathrm{af,typeI}} \,, \tag{7}$$

$$e_{\uparrow\downarrow}^{\text{af,typeII}} = \frac{1}{3} \cdot E_{\text{af,typeII}} - e_{\uparrow\uparrow}^{\text{af,typeII}} \ . \tag{8}$$

In general, $e_{\uparrow\uparrow}^{\text{fm}}$, $e_{\uparrow\uparrow}^{\text{af,typeI}}$, and $e_{\uparrow\uparrow}^{\text{af,typeII}}$ have different values because they have different local magnetic moments (see Sec. 3.1). In order to estimate $e_{\uparrow\uparrow}^{\text{af,typeI}}$ and $e_{\uparrow\uparrow}^{\text{af,typeII}}$ from $e_{\uparrow\uparrow}^{\text{fm}}$, which is determined through Eq. (4), a relation between the pair interaction energy and the magnetic moments is invoked in the mold of the Heisenberg model. In the Heisenberg model, the pair interaction energy of atoms is related to spin-angular momentum through[18, 19]

$$e_{ij} = -2J_{ij}\boldsymbol{S_i} \cdot \boldsymbol{S_j} \ , \tag{9}$$

where $J_{ij}$ is an exchange integral and $\boldsymbol{S_i}$ and $\boldsymbol{S_j}$ are a total spin-angular momentum. Using the relation, $\boldsymbol{\mu_i} = -2\mu_B \boldsymbol{S_i}$ (where $\boldsymbol{\mu_i}$ is the magnetic moment and $\mu_B$ is the Bohr magneton), Eq. (9) is converted into

$$e_{ij} = -2J_{ij} \frac{\boldsymbol{\mu_i} \cdot \boldsymbol{\mu_j}}{4\mu_B^2} \ . \tag{10}$$

Once $J_{\uparrow\uparrow}$ (or $J_{\uparrow\downarrow}$) are determined, this equation allows us to calculate the pair interaction energy between any magnitudes of local magnetic moments. The value of $J_{\uparrow\uparrow}$ is derived from Eqs. (4) and (10) using the local magnetic moment of up-spin in the ferromagnetic state, $\mu_{\uparrow}^{\text{fm}}$, which is obtained from the band calculation. On the other hand, the value of $J_{\uparrow\downarrow}^{\text{typeI}}$ (or $J_{\uparrow\downarrow}^{\text{typeII}}$) is estimated from the local magnetic moments of up- and down-spins in the antiferromagnetic state, $\mu_{\uparrow}^{\text{af,typeI}}$ and $\mu_{\downarrow}^{\text{af,typeI}}$ (or $\mu_{\uparrow}^{\text{af,typeII}}$ and $\mu_{\downarrow}^{\text{af,typeII}}$) with Eq. (10) by combining with Eq. (7) (or Eq. (8)), where $e_{\uparrow\uparrow}^{\text{af,typeI}}$ (or $e_{\uparrow\uparrow}^{\text{af,typeII}}$) can be calculated using $J_{\uparrow\uparrow}$ estimated above. As stated above, Eq. (10) makes it possible to set any magnitudes of local magnetic moments. In the present study, therefore, seven different local magnetic moments are assigned: 0.6, 0.4, 0.2, 0.0, –0.2, –0.4, and –0.6 $\mu_B$.

It is noteworthy that although the relation between the pair interaction energy and the local magnetic moment is derived by considering only exchange interaction energy in Eq. (10), the relation should be determined by taking into account both contributions of exchange energy and the kinetic energy of electrons.[18] As shown in Fig. 2, when only an exchange interaction energy is considered, the total energy monotonically decreases with the magnitude of the local magnetic moment. This means that if bigger local magnetic moments are assigned, a system prefers larger local magnetic moments than those at the ground state. On the other hand, when the kinetic energy of electrons is included, the total energy would have a minimum point at a given magnitude of the magnetic moment (which corresponds to the one at the ground state). However, the inclusion of the kinetic energy of electrons is beyond the scope of this work, and Eq. (10) is used here by setting the maximum magnitude of the local magnetic moments as 0.6 $\mu_B$, which is close to that at the ground state (see Sec. 3.1). Furthermore, the interval between the local magnetic moments specified above is arbitrarily chosen in this preliminary study. Although a more detailed analysis is required to examine the effect of the interval (or the number of specified local magnetic moments), that too is beyond the present scope.

## 2.3. Entropy and free energy

The configurational entropy is formulated within the tetrahedron approximation[20] of CVM and is given as

$$S = k_B N \left[ 6 \sum_{i,j} L(y_{ij}) - 5 \sum_{i} L(x_i) - 2 \sum_{i,j,k,l} L(w_{ijkl}) + 1 \right], \tag{11}$$

where $N$ is the total number of lattice points; $k_B$ is the Boltzmann constant; $x_i$, $y_{ij}$, and $w_{ijkl}$ are cluster probabilities of the point, pair, and tetrahedron, respectively; $i, j, k,$ and $l$ specify the local magnetic moments of Ni; and $L(x) \equiv x \ln x - x$. By viewing the various local magnetic moments as different atomic species, the calculation here is treated as the one in a seven-component alloy system. In alloys, the tetrahedron approximation is regarded as the minimum meaningful approximation that provides a combination of reasonable accuracy and acceptable computational burden.[21]

Together with the internal energy term given by Eq. (2), the final expression for the Helmholtz free energy becomes

$$F = \frac{N \cdot Z}{2} \sum_{i,j} e_{ij} y_{ij} - k_B T N \left[ 6 \sum_{i,j} L(y_{ij}) - 5 \sum_{i} L(x_i) - 2 \sum_{i,j,k,l} L(w_{ijkl}) - 1 \right]. \quad (12)$$

The equilibrium state can be determined by minimizing Eq. (12) with respect to the tetrahedron cluster probabilities, $w_{ijkl}$, and the interatomic distance, $r$ ;

$$\left. \frac{\partial F}{\partial w_{ijkl}} \right|_{r,T} = 0 \quad (13)$$

and

$$\left. \frac{\partial F}{\partial r} \right|_{x_i, y_{ij}, w_{ijkl}, T} = 0 \,. \quad (14)$$

The actual minimization is carried out by the Natural Iteration Method.[20)]

## 3. Results and Discussion

### 3.1. Total energy

The total energies and the magnetic moments in nonmagnetic, ferromagnetic, and two antiferromagnetic states calculated by VASP are shown in Figs. 3 and 4, respectively. The ferromagnetic and antiferromagnetic states have lower energies than the nonmagnetic state, and the fact that the lowest energy is attained at the ferromagnetic state corresponds to the experimental observation that Ni is ferromagnetic. The magnitudes of local magnetic moments in all states become bigger as the lattice expands. This represents magneto-volume effects.[22)]

### 3.2. Pair interaction energy

Pair interaction energies, *u*–*u* (or $e_{\uparrow\uparrow}$) and *u*–*d* (or $e_{\uparrow\downarrow}$), are shown in Figs. 5–7, respectively. There are two pair interaction energies in $e_{\uparrow\downarrow}$ depending on which antiferromagnetic energy, type-I or type-II, is used. In these figures, pair interaction energies between up-spin (or down-spin) and non-spin, *u*–*n* (or *d*–*n*), are not included in the insets because they are assumed to be the same as the pair energy of *n*–*n*. Since the pair interaction energies for *u–u* are negative and those for *u–d* are positive with respect to *n–n* pairs, it indicates that a ferromagnetic state is stabilized at low temperatures.

There is a difference between pair energies in 'up-spin and down-spin' extracted from the two different antiferromagnetic states, type-I and type-II. This implies that interactions between local magnetic moments are effective over a longer range than that of the first-nearest-neighbors. Thus, it would be important to employ longer magnetic interaction energies in order to estimate the energy term more reliably.

### 3.3. Curie temperature

Figure 8 shows the temperature dependence of the long-range-order (LRO) parameters, $\eta$, calculated in this work. The LRO parameters are defined as

$$\eta = \frac{|x_{up} - x_{down}|}{x_{up} + x_{down}}\ , \tag{15}$$

where $x_{up}$ is the sum of the point-cluster probabilities of up-spins (i.e., their local magnetic moments are 0.6, 0.4, and 0.2 $\mu_B$), and $x_{down}$ is the corresponding value with down-spins (i.e., their local magnetic moments are –0.2, –0.4, and –0.6 $\mu_B$). Whereas the LRO parameters are nearly 1.0 at low temperatures, they continuously approach zero as the temperature increases, and become zero at around 380 K for type-I and 425 K for type-II. This indicates that the magnetic state changes from ferromagnetic to paramagnetic at these temperatures, which correspond to the Curie temperature. However, the estimated Curie temperatures in this work, 380 K (for type-I) and 425 K (for type-II), are well below the experimental value of 627.2 K.[19]

Point-cluster probabilities of non-spin, up-spin, and down-spin are shown in Fig. 9. In this figure, it can be seen that a magnetic transition takes place as a result of cancellation between up-spins and down-spins, while the local magnetic moments do not entirely vanish. This suggests that this magnetic transition is from a ferromagnetic state to a paramagnetic state (not from a ferromagnetic to a nonmagnetic state).

The temperature dependence of the average squared magnitude of magnetic moments is shown in Fig. 10. It shows that the average magnitude of local magnetic moments decreases with temperature. It is known that whereas the absolute values of the magnetic moments do not change significantly in Fe and Co, they are sensitive to temperature in Ni.[23,24] The results shown in Fig. 10 agree with this observation.

There are some possible reasons for the discrepancy between the calculated Curie temperature and the experimental data. One of the main reasons can be ascribed to the fact that spin configurations whose local magnetic moments are aligned in the <001> direction are used in the total-energy calculations. It is, however, known that the preferable spin direction in Ni is the <111> direction. Therefore, if the total energies with spin alignments in <111> directions are used, the Curie temperature would be increased; this will be the subject of future calculations. In addition, the underestimation of the Curie temperature may be ascribed to the use of the first-nearest-neighbor model, where only the nearest neighbor pair interactions are taken into account.

## 4. Conclusions

We have calculated the Curie temperature of pure Ni by using a modified CVM in which the magnetic freedom is converted into the configurational freedom of a multicomponent alloy system. The energetics involved in the calculations are evaluated from first-principles total-energy calculations for Ni, where some different spin configurations are considered.

In this preliminary study, seven different magnitudes of local magnetic moments are used. The calculated Curie temperatures, 380 K and 425 K, are underestimated compared with the experimental data, 672.2 K. There are some possible reasons for this discrepancy. One of the main reasons is that spin alignments in <001> directions are assumed in a band calculation, even though it is known that the preferable spin direction of Ni is the <111> direction.

In terms of the configurational freedom of spin alignments, the present work focuses exclusively on a collinear configuration. An extension of the present study to a noncollinear spin configuration should be straightforward for the entropy term by following the same formalism of CDCVM with an extended number of species. However, difficulties might arise in energy calculations to distinguish very small energy differences in spin configurations. Furthermore, the above points have a common deficiency in terms of the wide range of interactions involved. In fact, the present calculations are limited to the first-nearest-neighbor pair interaction energies in order to focus on the extension of the idea in the CDCVM (i.e., a conversion from a freedom of local displacements into the configurational one) to magnetic freedoms. For a more reliable estimation of the energy term, a careful electronic-structure calculation would be required.

## Acknowledgement

This paper is based on results obtained from a project commissioned by the New Energy and Industrial Technology Development Organization (NEDO).

References

1) R. Kikuchi: Phys. Rev. **81** (1951) 988–1003.
2) D. de Fontaine and R. Kikuchi: Fundamental Calculations of Coherent Phase Diagrams, In: Application of Phase Diagrams in Metallurgy and Ceramics: Proceeding of a Workshop Held at the National Bureau of Standards, Gaithersburg, Maryland, January 10–12, 1977, G. C. Carter (Ed.), US Department of Commerce: Washington DC (1978); Vol. 2., pp. 999–1026.
3) P. Cenedese and R. Kikuchi: Physica A (Amsterdam, Neth.) **205** (1994) 747–755.
4) R. Kikuchi: J. Chem. Phys. **47** (1967) 1664–1668.
5) T. Morita and T. Tanaka: Phys. Rev. **145** (1966) 288–295.
6) C. G. Schön and G. Inden: J. Magn. Magn. Mater. **232** (2001) 209–220.
7) Y. Chen, S. Iwata, and T. Mohri: Rare Met. (Beijing, China) **25** (2006) 437–440.
8) T. Mohri and Y. Chen: Mater. Trans. **45** (2004) 1478–1484.
9) R. Kikuchi: J. Phase Equilib. **19** (1998) 412–421.
10) T. Mohri: JOM **65** (2013) 1510–1522.
11) T. Mohri: Comput. Mater. Sci. **49** (2010) S181–S186.
12) T. Mohri, Y. Chen, and N. Kiyokane: J. Alloys Compd. **577** (2013) S123–S126.
13) N. Kiyokane and T. Mohri: Philosophical Magazine **98** (2018) 1005–1017.
14) G. Kresse and J. Furthmüller: Phys. Rev. B **54** (1996) 11169–11186.
15) J. P. Perdew, K. Burke, and M. Ernzerhof: Phys. Rev. Lett. **77** (1996) 3865–3868.
16) J. P. Perdew and Y. Wang: Phys. Rev. B **45** (1992) 13244–13249.
17) L. A. Girifalco: Statistical mechanics of solids, OUP USA (2003).
18) M. Shiga: Introduction to Magnetism, Uchida Rokakuho Publishing: Tokyo (2007) (in Japanese).
19) C. Kittel: Introduction to Solid State Physics, Wiley: New York (1986), 6th ed.
20) R. Kikuchi: J. Chem. Phys. **60** (1974) 1071–1080.
21) T. Mohri: Metallurgical and Materials Transactions A 48 (2017) 2753–2770.
22) Y. Takahashi: Spin Fluctuation Theory of Itinerant Electron Magnetism, Springer: Heidelberg (2013), pp. 131–177.
23) M. Pajda, J. Kudrnovský, I. Turek, V. Drchal, and P. Bruno: Physical Review B 64.17 (2001) 174402.
24) A. V. Ruban, S. Khmelevskyi, P. Mohn, and B. Johansson: Physical Review B 75.5 (2007) 054402.

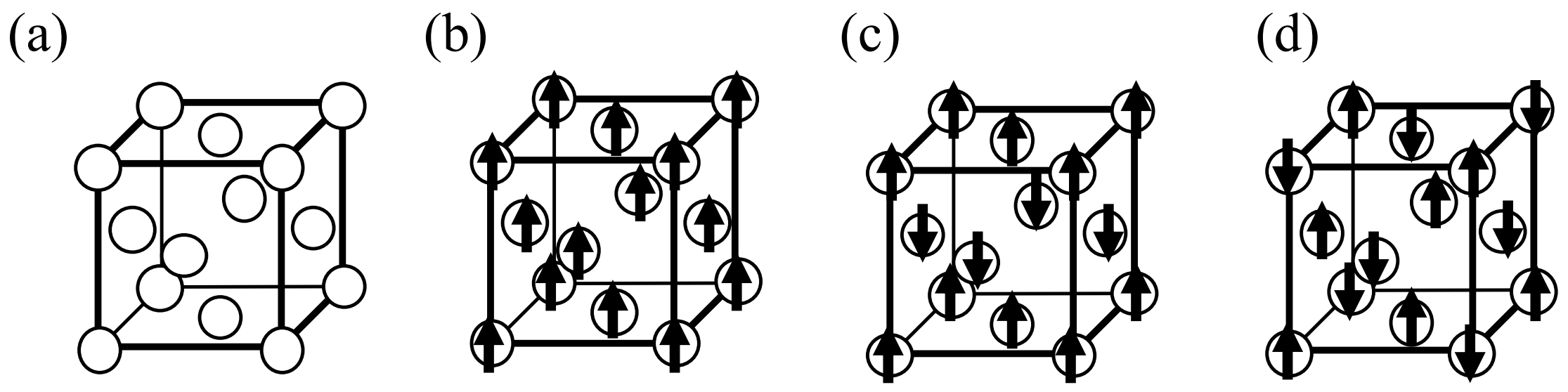

Fig. 1. Various spin configurations; (a) nonmagnetic (nm) state, (b) ferromagnetic (fm) state, and (c) and (d) antiferromagnetic (af) states. (c) and (d) are, respectively, called type-I and type-II.

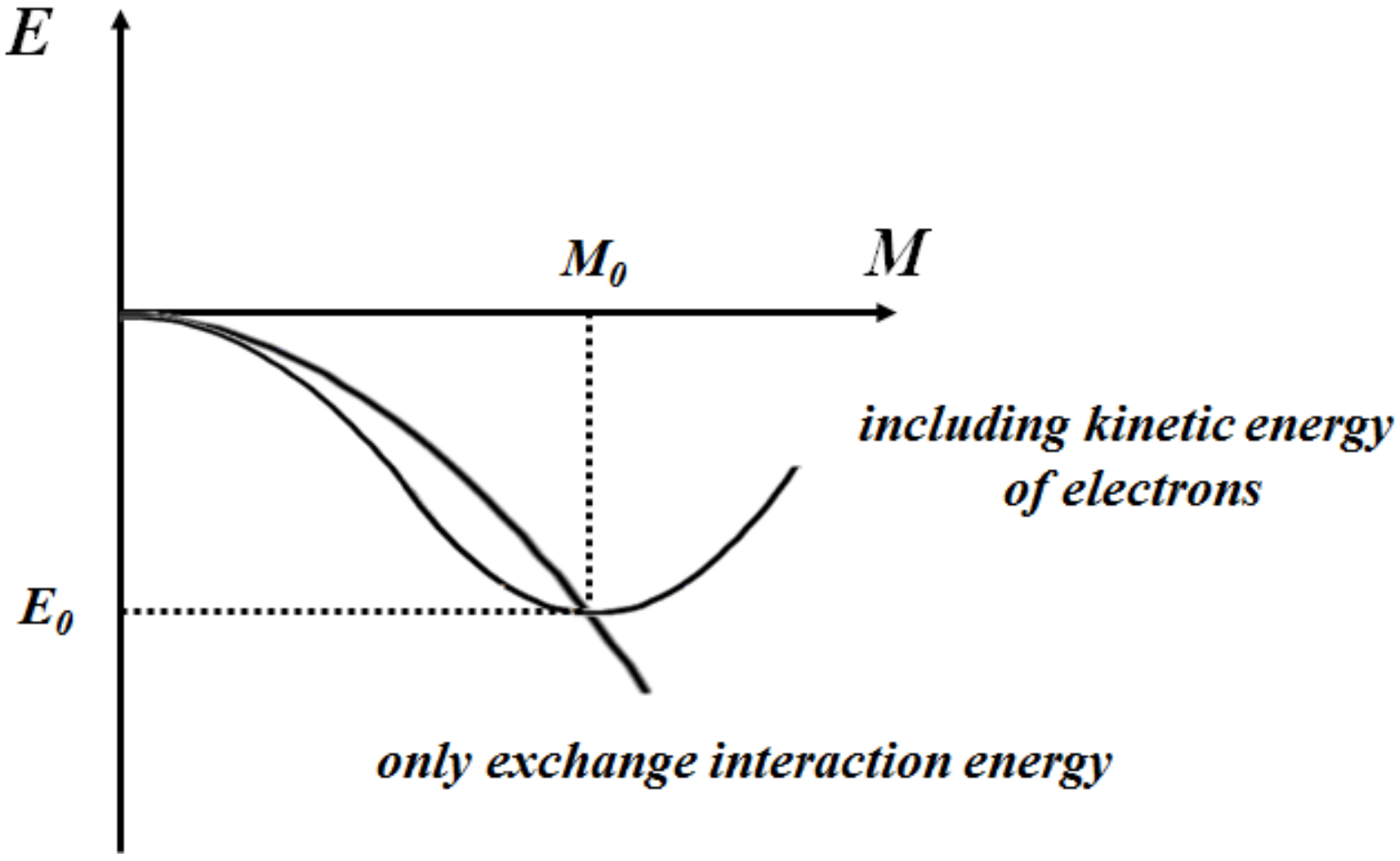

Fig. 2. Schematic explanation of the difference between energies with and without the kinetic energy of electrons. Horizontal line indicates the magnitude of local magnetic moments. $E_0$ represents the minimum energy when the kinetic energy of electrons is included ($M = M_0$).

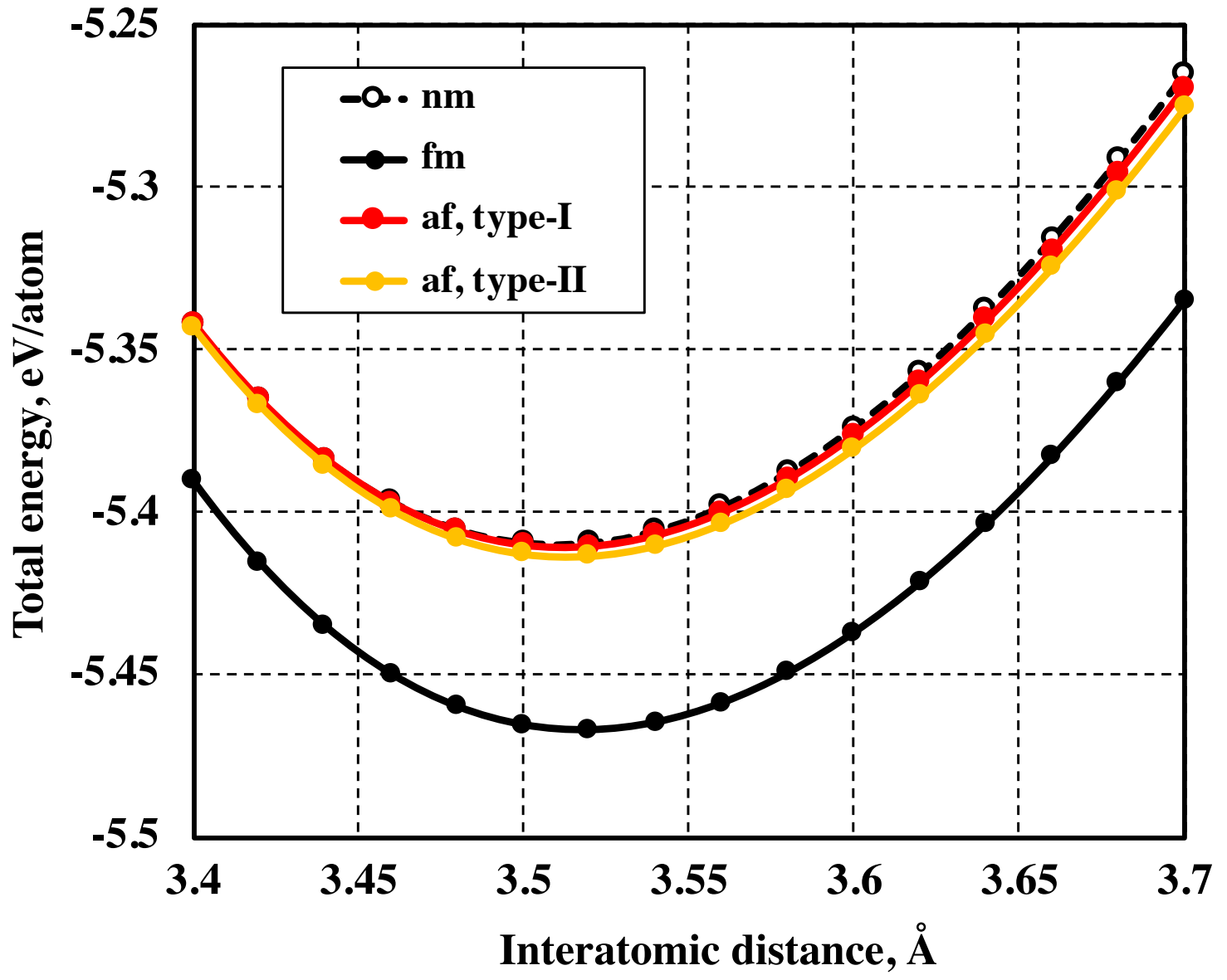

Fig. 3. Total energy of nonmagnetic (nm), ferromagnetic (fm), and two antiferromagnetic (type-I and type-II) states as a function of interatomic distance.

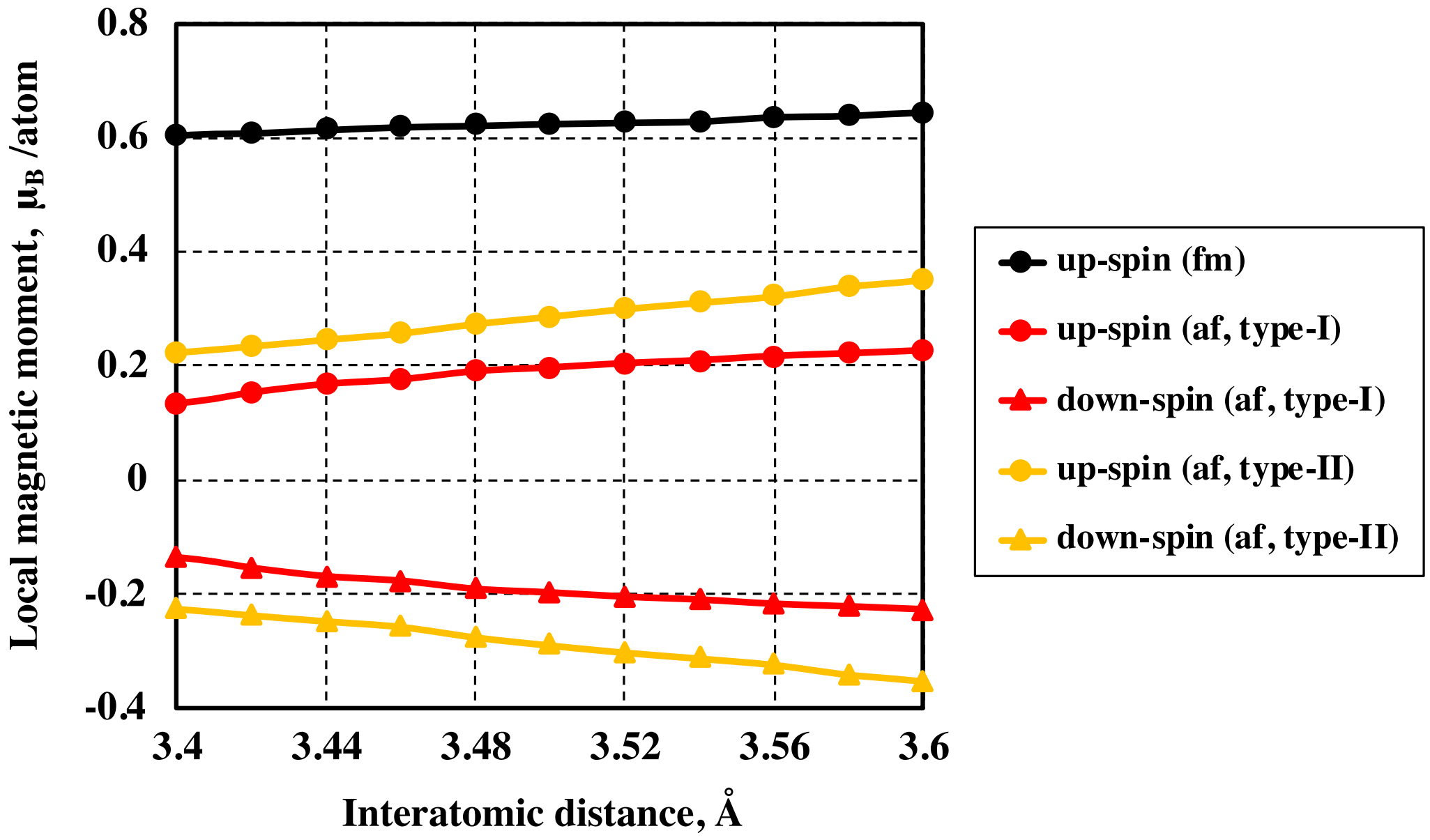

Fig. 4. Magnitude of local magnetic moments at each atom. The circles/triangles are the average of up-/down-spins.

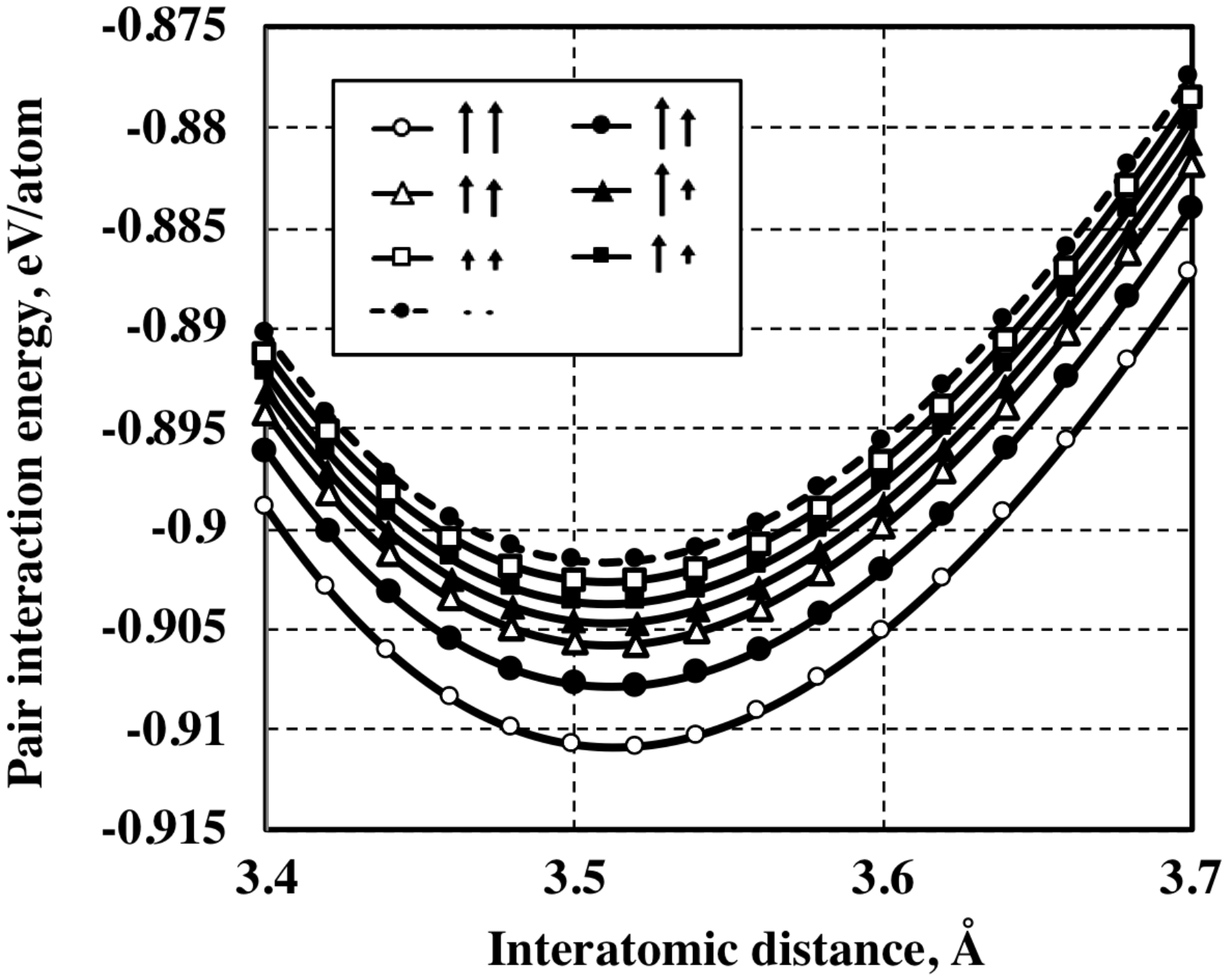

Fig. 5. Pair interaction energy between *u*-*u* pairs. Open/solid marks are those with same/different local magnetic moments. The broken line is the energy of the *n*-*n* pair. The *u*-*n* pairs are assumed to be the same as the *n*-*n* pair and are not shown in the inset (because it becomes tedious).

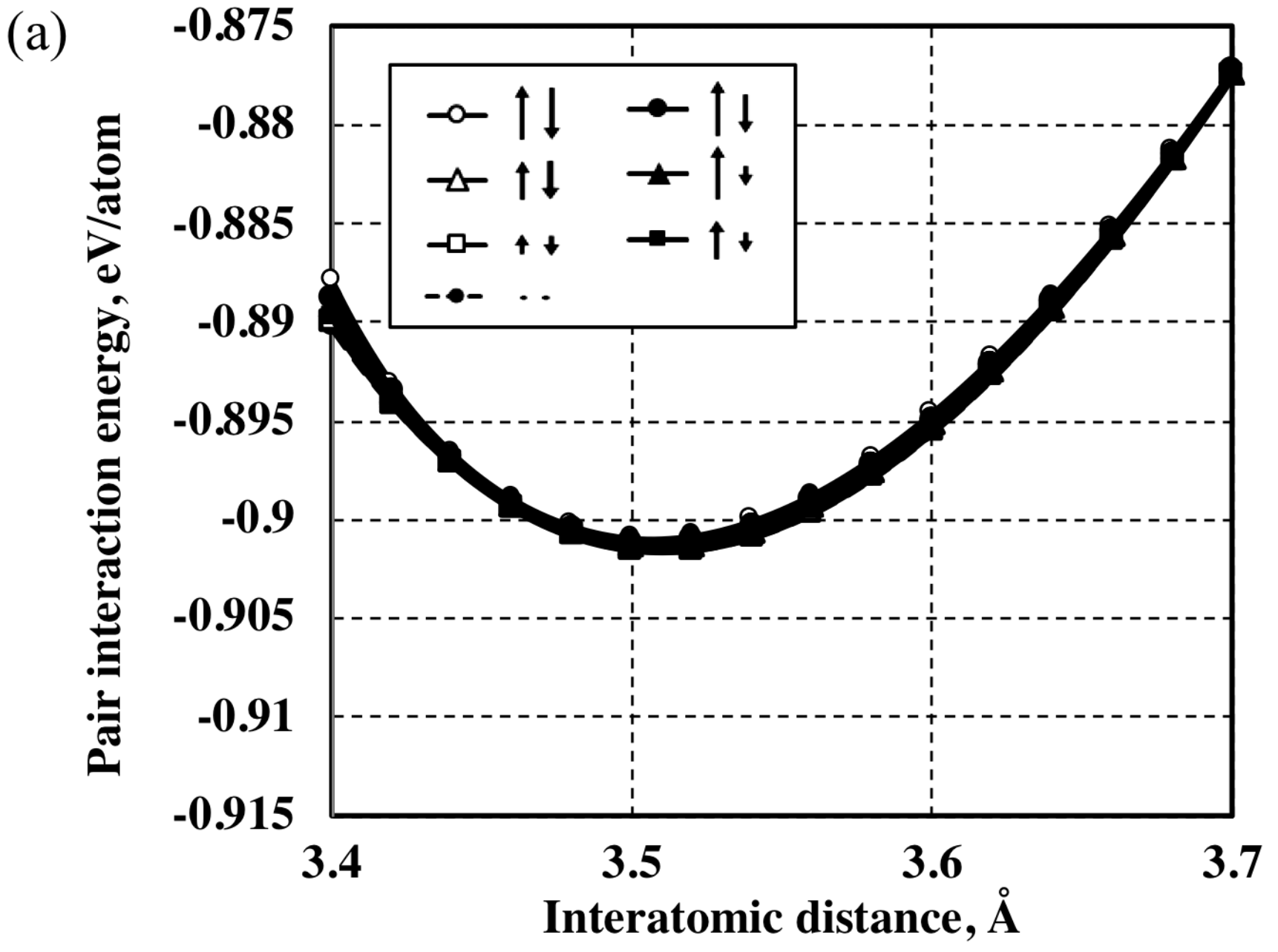

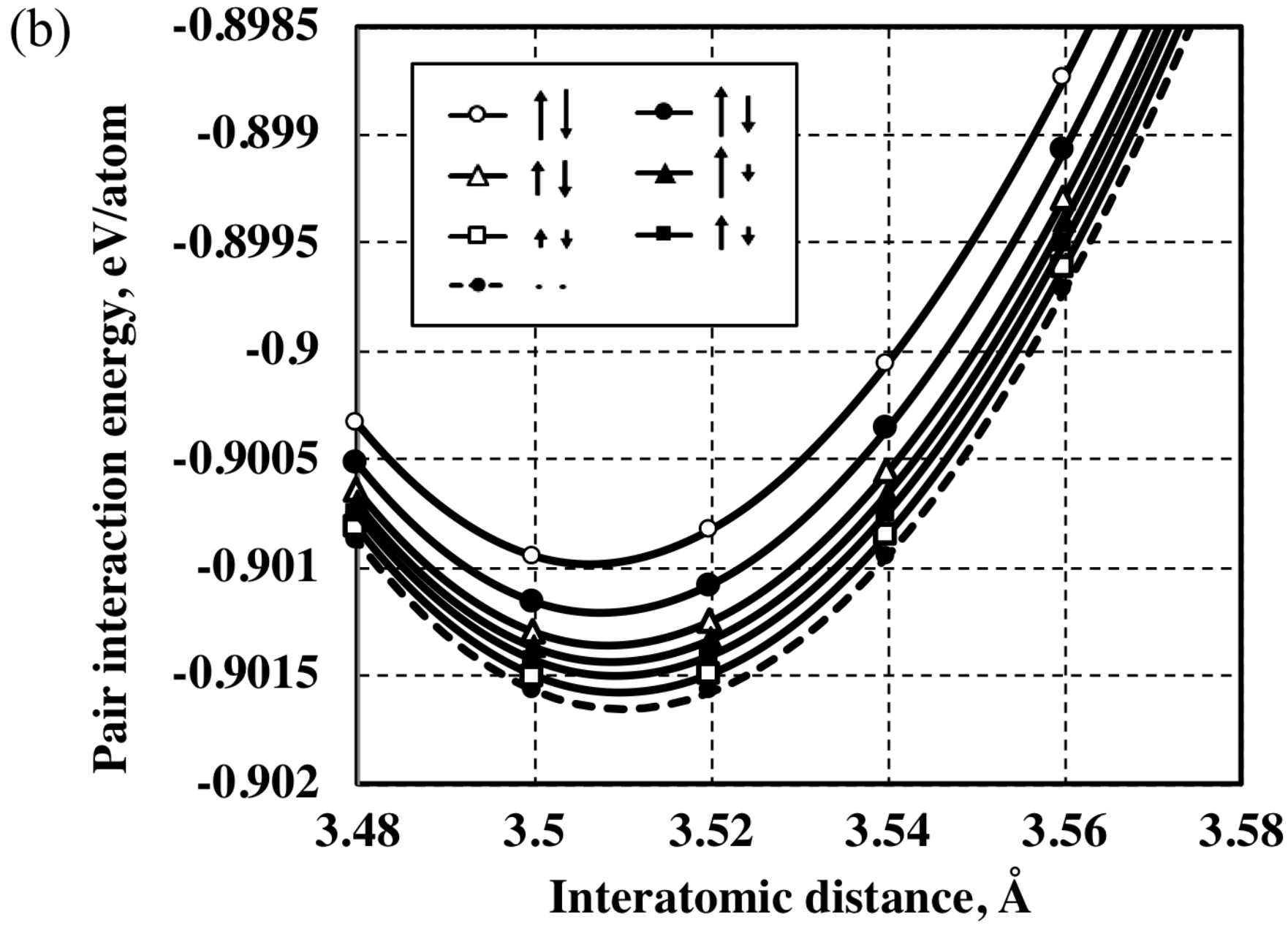

Fig. 6. Pair interaction energy between *u-d* pairs for type-I. (b) is a magnification of (a). Open/solid marks indicate those with same/different local magnetic moments. The broken line is the energy of the *n-n* pair. The *d-n* pairs are assumed to be the same as the *n-n* pair and are not shown in the inset (because it becomes tedious).

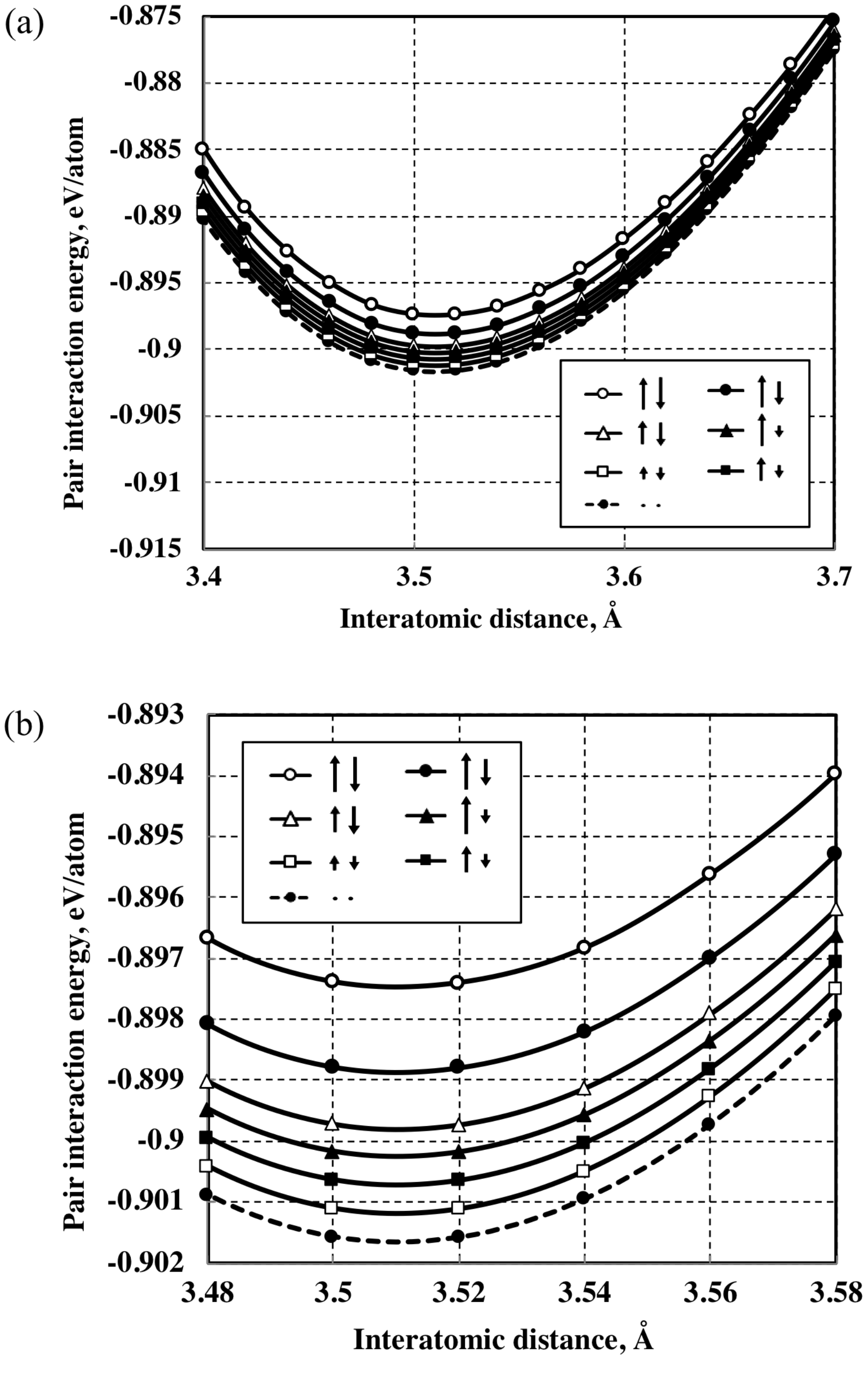

Fig. 7. Pair interaction energy between *u*-*d* pairs for type-II. (b) is a magnification of (a). Open/solid marks indicate those with same/different local magnetic moments. The broken line is the energy of the *n*-*n* pair. The *d*-*n* pairs are assumed to be the same as the *n*-*n* pair and are not shown in the inset (because it becomes tedious).

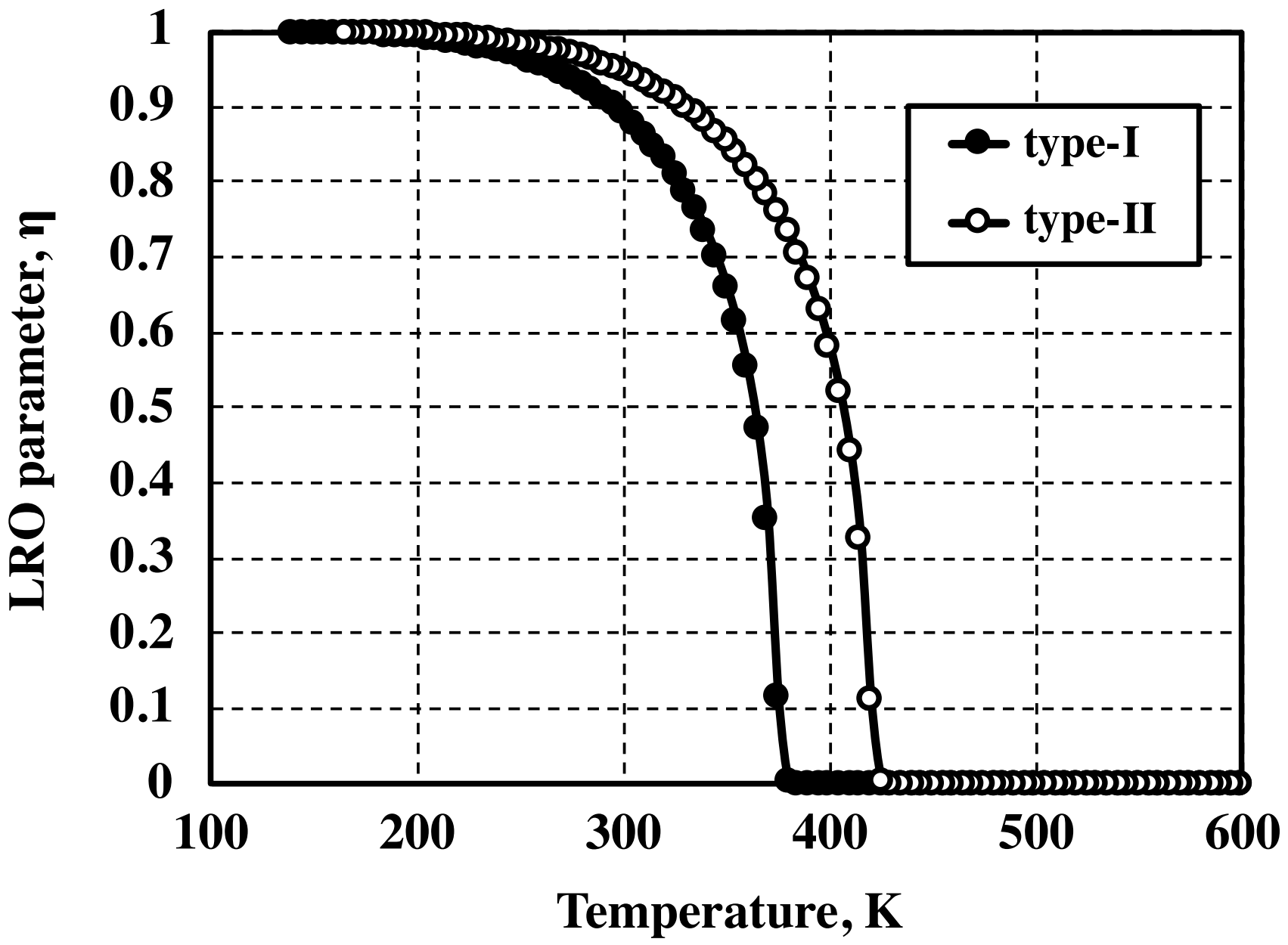

Fig. 8. Temperature dependence of LRO parameter. Solid and open circles indicate the results obtained from different antiferromagnetic spin configurations, type-I and type-II.

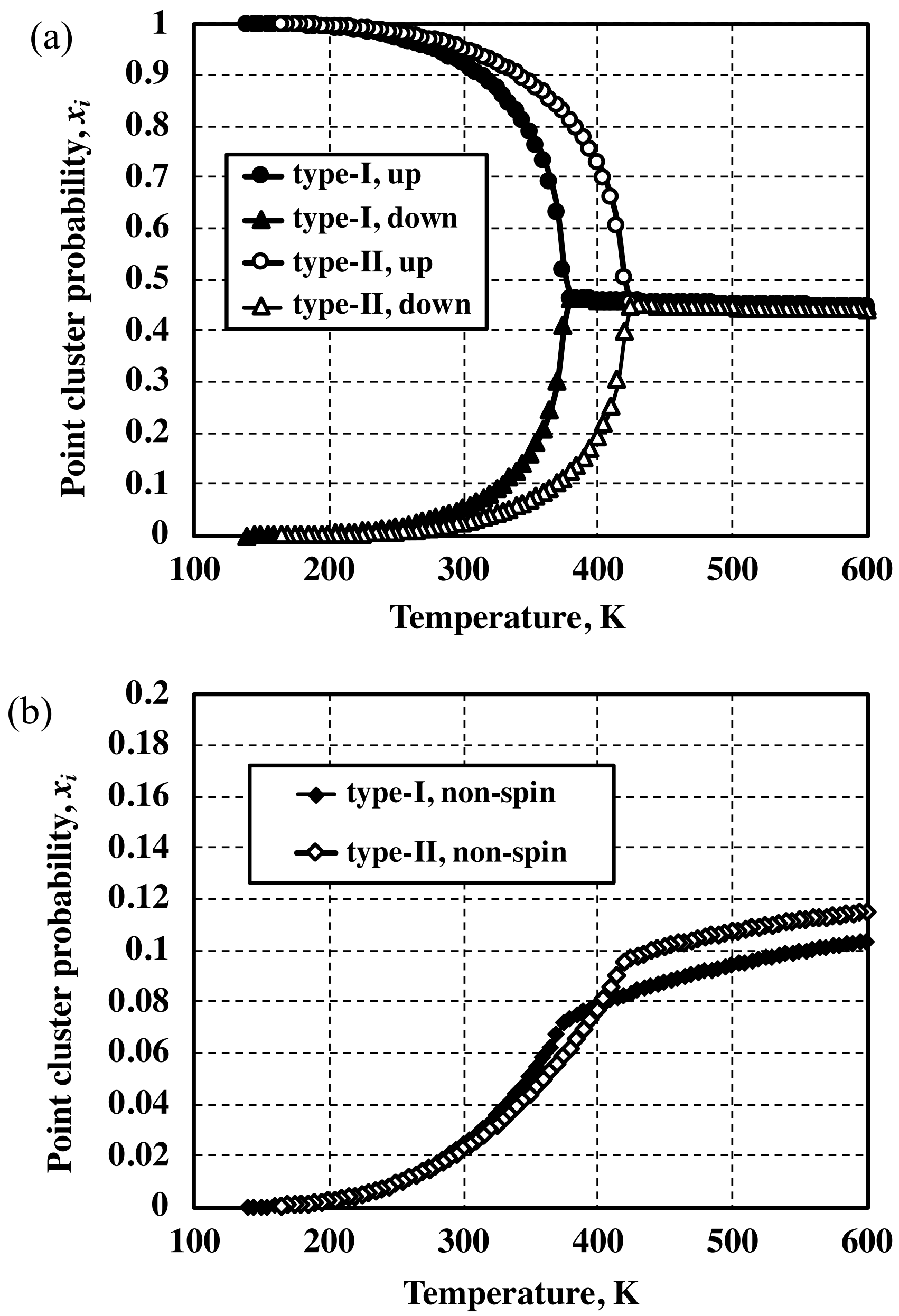

Fig. 9. Temperature dependence of point cluster probabilities of up-, down-, and non-spins.

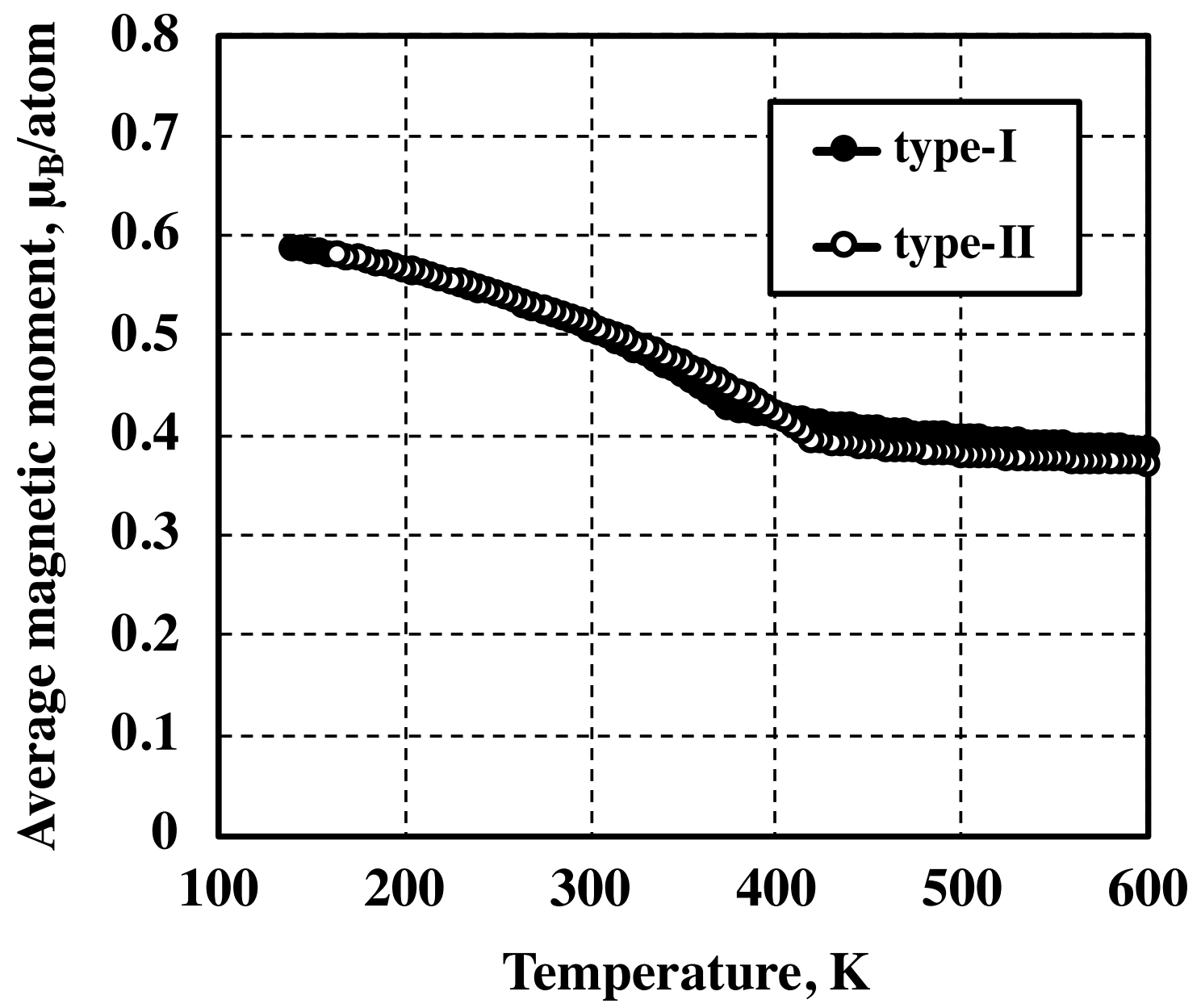

Fig. 10. Temperature dependence of average magnitudes of absolute values of magnetic moments. Solid and open circles indicate the results obtained from different antiferromagnetic spin configurations, type-I and type-II.